\documentclass[12pt]{iopart}
\usepackage{graphicx,verbatim}
\usepackage{braket}
\usepackage{color}
\usepackage{stix}
\usepackage{soul}
\setstcolor{blue}
\usepackage[normalem]{ulem}
\usepackage[x11names]{xcolor}
\usepackage{oplotsymbl}
\usepackage[group-separator={,}]{siunitx}

\expandafter\let\csname equation*\endcsname\relax
\expandafter\let\csname endequation*\endcsname\relax
\usepackage{amsmath}

\usepackage[square,sort&compress,numbers]{natbib}
\usepackage[colorlinks,allcolors=blue]{hyperref} 
\usepackage[capitalise,noabbrev,nameinlink]{cleveref}

\begin{document}

\title[Determinant- and Derivative-Free QMC within the SRW]{Determinant- and Derivative-Free Quantum Monte Carlo Within
the Stochastic Representation of Wavefunctions}
\author{Liam Bernheimer$^{1}$, Hristiana Atanasova$^{1}$, Guy Cohen$^{1,2}$}

\address{$^{1}$ School of Chemistry, Tel Aviv University, Tel Aviv 6997801, Israel}
\address{$^{2}$ The Raymond and Beverley Sackler Center for Computational Molecular and Materials Science, Tel Aviv University, Tel Aviv 6997801, Israel}
\ead{gcohen@tau.ac.il}

\date{\today}

\begin{abstract}
    Describing the ground states of continuous, real-space quantum many-body systems, like atoms and molecules, is a significant computational challenge with applications throughout the physical sciences.
    Recent progress was made by variational methods based on machine learning (ML) ansatzes.
    However, since these approaches are based on energy minimization, ansatzes must be twice differentiable. This (a) precludes the use of many powerful classes of ML models; and (b) makes the enforcement of bosonic, fermionic, and other symmetries costly.
    Furthermore, (c) the optimization procedure is often unstable unless it is done by imaginary time propagation, which is often impractically expensive in modern ML models with many parameters.
    The stochastic representation of wavefunctions (SRW), introduced in Nat Commun 14, 3601 (2023), is a recent approach to overcoming (c).
    SRW enables imaginary time propagation at scale, and makes some headway towards the solution of problem (b), but remains limited by problem (a).
    Here, we argue that combining SRW with path integral techniques leads to a new formulation that overcomes all three problems simultaneously.
    As a demonstration, we apply the approach to generalized ``Hooke's atoms'': interacting particles in harmonic wells.
    We benchmark our results against state-of-the-art data where possible, and use it to investigate the crossover between the Fermi liquid and the Wigner molecule within closed-shell systems.
    Our results shed new light on the competition between interaction-driven symmetry breaking and kinetic-energy-driven delocalization.
\end{abstract}
\maketitle

\section{Introduction} \label{intro}
The static properties of molecules and materials at low temperatures can be obtained from the ground state of the Schr\"odinger equation describing them.
However, exact expressions for the ground state are available for only a few special cases where the degrees of freedom are either very few, effectively captured by a mean-field theory, or obey a highly specialized symmetry condition.
One of the most successful and enduring paradigms for going beyond these limits are variational Monte Carlo (VMC) approaches, where many-parameter wavefunction ansatzes with bosonic \cite{PhysRev.138.A442} or fermionic \cite{PhysRevB.16.3081} exchange symmetry are optimized with respect to energy by way of stochastic integration.
The quality of the approximation depends on the expressiveness of the ansatz and on the optimization method, but in practice provides some of the highest accuracies available in quantum chemical calculations \cite{morales_multideterminant_2012}.

The recent surge of machine learning (ML) technologies and neural networks (NNs) in particular led to their use as ansatzes within VMC, which have rapidly attained state-of-the-art performance for several types of systems \cite{taddeiIterativeBackflowRenormalization2015,ruggeri_nonlinear_2018,Manzhos2020,hermann_deep-neural-network_2020,ferminet,spencer_better_2020,wilson_simulations_2021,klus_symmetric_2021,schatzle_convergence_2021,	keith_combining_2021,xie_ab_initio_2022,liInitioCalculationReal2022a,schatzle_deepqmc_2023,von_glehn_self-attention_2023,wilsonNeuralNetworkAnsatz2023,pescia_message-passing_2023,ren_towards_2023,kim_neural-network_2023,lange_architectures_2024,aldossary_silico_2024}.
Here, we will focus on implicit real-space, first-quantized representations of wavefunctions, where no single-particle orbital basis need be defined and no overlap integrals need be calculated.
Approaches of this kind often leverage physical or chemical intuition to make the solution more efficient and accurate \cite{hermann_deep-neural-network_2020}, but we will also focus on relatively generic, and therefore somewhat simpler, ansatzes.

Implicit variational approaches have several distinct disadvantages that stand in the way of further progress.
First, they require the evaluation of the second derivative of the wavefunction with respect to coordinates.
Modern ML models are necessarily constructed so as to be once-differentiable with respect to their parameters, and the need to also be twice differentiable with respect to the inputs introduces additional constraints that may not be desirable.
For example, it precludes the use of highly efficient activation functions like the rectified linear unit (ReLU) \cite{agarap_deep_2019}, which has a piecewise-constant first derivative.

Second, a major trade-off of using first-quantized wavefunctions is that---at least for fermions, and preferably also for bosons---the exchange symmetry must be incorporated into the ansatz \cite{hutter_representing_2020}.
This has most commonly been achieved by using determinants \cite{ferminet,hermann_deep-neural-network_2020}, at a computational cost that scales as $\mathcal{O}(N^3)$ with the number of particles $N$.
An approach based on Vandermonde determinants reduces this to $\mathcal{O}(N^2)$ \cite{acevedo_vandermonde_2020,pang_on2_2022}, but with some constraints on universality \cite{richter-powell_sorting_2023}.
A recent sorting-based approach \cite{richter-powell_sorting_2023} reduces this to $\mathcal{O}(N\log{N})$ with the same constraints, and can lift those constraints at a cost of $\mathcal{O}(N^2\log{N})$.
Nevertheless, several questions regarding smoothness and universal representability remain open \cite{hutter_representing_2020,pang_on2_2022,acevedo_vandermonde_2020}.

Third, optimizing a wavefunction ansatz with respect to energy can be with the generalized gradient descent techniques typically employed in ML, which can fail to converge to a good approximation of the ground state.
Practitioners of VMC often rely on effectively performing imaginary time evolution using stochastic reconfiguration (SR) techniques, known as natural gradient descent in the ML literature \cite{PhysRevB.64.024512,PhysRevB.71.241103}.
However, this comes with a computational cost that is cubic in the number of parameters, making it nonviable for ML models with more than several thousand parameters.
The result is limited access to many modern ML architectures, which can contain millions or even trillions of parameters \cite{Birhane2023}.
A recently propose alternative shows that alternatively, SR can be carried out at a cost that is cubic with the number of samples, which can be lower in many circumstances \cite{rende_simple_2023,chenEmpoweringDeepNeural2024}.

Recently, we introduced the stochastic representation of wavefunction (SRW) \cite{Atanasova2023}, which replaces the (unsupervised) search for a minimum energy into a series of (supervised) regression steps.
The regression is performed with respect to a series of sampled points, for each of which the value of the wavefunction after a short imaginary time evolution has been calculated by applying an approximate propagator.
The idea of replacing energy optimization with propagation and regression had previously appeared in the literature for second-quantized and spin systems, where it has been called ``self-supervised learning'' \cite{jonsson_neural-network_2018,medvidovic_classical_2021,giuliani_learning_2023} and ``supervised  wavefunction optimization'' \cite{kochkov_variational_2018,kochkov_learning_2021,luo_autoregressive_2022,luo_gauge_2023}.
The idea has also been applied to real time evolution in such models \cite{schmitt_quantum_2020,gutierrez_real_2022,ledinauskas_scalable_2023,sinibaldi_unbiasing_2023}, and a related method uses the Lanczos algorithm to achieve energy minimization \cite{chen_systematic_2022}.
Our own work on SRW \cite{Atanasova2023} focused on first-quantized systems, and introduced a somewhat successful stochastic projection technique for enforcing exchange symmetry.
However, it did not circumvent the need for twice differentiable ansatzes because the propagator was enacted by applying the Hamiltonian operator to the wavefunction.

Here, we show that performing the imaginary time evolution step within SRW using Feynman path integrals leads to a new framework that is more general than variational methods.
In particular, the framework we present removes the need for differentiable or even continuous ansatzes, while still providing well-defined non-variational energy estimates and eventually a variational upper bound.
This not only allows for the use of more efficient ML ansatzes, but also enables enforcement of exchange symmetry at a cost of $\mathcal{O}(N\log{N})$ without sacrificing universal representability.

To establish the new method, we apply it on a generalized Hooke's atom \cite{hookesPhysRev.128.2687}, where instead of electrons and a nucleus we have several fermions or bosons in a harmonic trap with Coulomb interaction.
This toy model is used to investigate electrons in quantum dots, but is also a commonly used benchmark for quantum Monte Carlo methods, to which we present some comparisons.
For fermions in particular, our method allows us to access low energy states at a large variety of interaction strengths all the way from the degenerate noninteracting quantum state to the semiclassical Wigner molecule, which exhibits spontaneous symmetry breaking.
We investigate the density profile in two spin states, demonstrating that several closely competing states with different symmetries are possible.

\section{Methods}
We propose a general SRW-based approach to finding an estimate for the ground state wavefunction of a given system, focusing on the first-quantized description of $N$ particles in a continuous $d$-dimensional space.
A loose outline of the procedure follows.
In the sections below, each step will be discussed in detail.
\begin{enumerate}
\item
An initial guess for the wavefunction, $\psi\left(\mathbf{R}\right)$, is provided; here $\mathbf{R}=\left( \mathbf{r}_{1},\ldots,\mathbf{r}_{N}\right)$ denotes the $Nd$ coordinates of all particles in space.
The guess may, for example, come from the noninteracting limit, a mean-field approximation, or any other approximate scheme that produces a wavefunction; but can also be a random neural network.

\item
To propagate the wavefunction by an imaginary time step, a collection of $M$ points $\mathbf{R}_m$ in coordinate space is selected.
These sampled points must be chosen so as to describe the state sufficiently well.

\item
The propagated wavefunction at each point, $e^{-\hat{H}\tau/\hbar}\psi\left(\mathbf{R}_{m}\right)$, is obtained by path integration as explained in subsections \ref{PIsection} and \ref{mcseciton}.
The procedure of subsection \ref{enrgsection} is employed to obtain a nonvariational estimate of the energy at essentially no additional cost.

\item
The sampled points and wavefunctions are used to find a new estimate $\psi\left(\mathbf{R}\right)$ for the propagated wavefunction by regression, using a machine learning model incorporating the explicit symmetry enforcement scheme from subsection \ref{symsection}, and constructed in accordance with subsection \ref{regsection}.

\item
Steps 2--4 are iterated until the energy estimator converges.

\item Finally, if desired, a variational upper bound for the energy can be obtained, as discussed in section \ref{enrgsection}.

\end{enumerate}

This path-integral-based SRW procedure is the main methodological result of this paper.
It is at first glance very similar to VMC techniques, but differs from it in several ways.
Even without path integrals, the SRW method overcomes convergence issues that plague VMC methods at scale, as previously discussed in the literature \cite{Atanasova2023}.
With path integrals at its core, however, it also implies that wavefunction ansatzes no longer need to be differentiable or even continuous.
As we discuss in detail below, this has major implications.

\subsection{Imaginary time propagation}
\label{PIsection}
\paragraph*{General considerations.}
Given a Hamiltonian $\hat{H}$ and a set of boundary conditions, the time-independent Schrödinger equation in the position basis takes the form
\begin{equation}
    \hat{H}\varphi_{n}\left(\mathbf{R}\right)=E_{n}\varphi_{n}\left(\mathbf{R}\right),
\label{Schrödinger}
\end{equation}
where $\varphi_{n}$ are the system's stationary states with corresponding energies $E_{n}$.
Any wavefunction or state obeying the same boundary conditions can formally be written in the energy representation, in terms of a set of coefficients  $c_{n}$:
\begin{equation}
	\psi\left(\mathbf{R}\right)=\sum_{n=0}^{\infty}c_{n}\varphi_{n}(\mathbf{R}).
	\label{eigendecomposition}
\end{equation}
If the ground state is non-degenerate, a state proportional to it can be found by starting from an arbitrary state with nonzero $c_0$ and propagating it to sufficiently large imaginary times:
\begin{equation}
\begin{aligned}\psi\left(\mathbf{R},\tau\right) &\equiv e^{-\hat{H}\tau/\hbar}\psi\left(\mathbf{R}\right) =\sum_{n=0}^{\infty}c_{n}\varphi_{n}(\mathbf{R})e^{-E_{n}\tau/\hbar}\\
 & \xrightarrow[E_{0}\leq E_{1}\leq\ldots\leq E_{n}]{\tau\rightarrow\infty} e^{-E_0\tau} \varphi_{0}(\mathbf{R}).
\end{aligned}
\label{time propagation}
\end{equation}
All corrections to this expression decay exponentially in imaginary time.
If the ground state is degenerate, the procedure described above will produce an arbitrary state in the manifold spanned by the set of lowest energy stationary states.

\paragraph*{Path integrals.}
One approach to performing time evolution in systems where the stationary states are intractable is path integration \cite{feynman_quantum_2010}.
If we divide the propagation time $\tau$ into $T$ equal intervals of length  $\varepsilon=\frac{\tau}{T}$, the exact result is given by the limit
\begin{equation}
\begin{aligned}
	e^{-\hat{H}\tau/\hbar}\psi\left(\mathbf{R}_{0}\right) & =\lim_{T\rightarrow\infty}\left(\frac{mT}{2\pi\tau\hbar}\right)^{\frac{TNd}{2}} \int_{\left(\mathbb{R}^{Nd}\right)^{T}} \mathrm{d}\mathbf{R}_{1}\cdots\mathrm{d}\mathbf{R}_{T}\\
	& \times\exp\left\{ -\frac{1}{\hbar}\sum_{j=1}^{T}\varepsilon\Biggl[\frac{m}{2}\left|\frac{\mathbf{R}_{j}-\mathbf{R}_{j-1}}{\varepsilon}\right|^{2}\right.\\
	&\qquad\qquad\qquad+V\left(\mathbf{R}_{j-1}\right)\Biggr]\Biggr\}\psi\left(\mathbf{R}_{T}\right),
\end{aligned}
\label{path integral}
\end{equation}
A numerical evaluation of such an expression is the starting point for path integral Monte Carlo (PIMC) techniques, but its evaluation at large imaginary times, and therefore large $T$, generically leads to sign problems \cite{ceperley_path_1995,foulkes_quantum_2001,carlson_quantum_2015}.

Let us focus on the substantially more modest goal of evaluating this over a small time interval $\Delta \tau$, and construct a minimal realization of the path integral.
Consider the $T=1$ case, i.e. $\Delta\tau=\varepsilon\rightarrow0$.
There,
\begin{equation}
	\begin{aligned}e^{-\Delta\tau\hat{H}/\hbar}\psi\left(\mathbf{R}_{0}\right) & \simeq\left(\frac{m}{2\pi\Delta\tau\hbar}\right)^{Nd/2} \times \int_{\mathbb{R}^{Nd}}\exp\Bigg\{\\
		& \left.-\frac{1}{\hbar}\Delta\tau\left[\frac{m}{2}\left|\frac{\mathbf{R}_{T}-\mathbf{R}_{0}}{\Delta\tau}\right|^{2}+V\left(\mathbf{R}_{0}\right)\right]\right\} \\
		& \times\psi\left(\mathbf{R}_{T}\right)\mathrm{d}\mathbf{R}_{T}.
	\end{aligned}
	\label{new path integral}
\end{equation}
This provides an estimate for the propagated wavefunction at the $\mathbf{R}_0$, in terms of an $Nd$-dimensional integral.
Interestingly, for any finite $\Delta \tau$, this expression contains no derivatives and remains well defined even for inputs that would destabilize the energy evaluation in VMC: $\psi\left(\mathbf{R}\right)$ can be neither differential nor continuous.

\paragraph*{Linear paths.}
Practically speaking, it will be useful to make $\Delta \tau$ as large as possible without sacrificing accuracy.
A straightforward improvement can be obtained by recognizing the exponent as an estimate for the Euclidean action for a constant velocity path, 
$\mathbf{R}^{L}\left(t\right)=\left(1-\frac{t}{\Delta\tau}\right)\mathbf{R}_{0}+\frac{t}{\Delta\tau}\mathbf{R}_{T}$,
with the potential treated as constant over the path.
A better approximation can then be obtained if we can efficiently integrate the potential along such a path.
The exact linear Euclidean action is given by
 \begin{equation}
		\begin{aligned}
		S_{\text{E}}^{\text{L}}(\mathbf{R}_{0},\mathbf{R}_{T},\Delta\tau) & =\int_{0}^{\Delta\tau}\left[\frac{m}{2}\left|\frac{\mathbf{R}_{T}-\mathbf{R}_{0}}{\Delta\tau}\right|^{2}+V\left(\mathbf{R}^{L}\left(t\right)\right)\right]\mathrm{d}t\\
		 & =\frac{m}{2\Delta\tau}\left|\mathbf{R}_{T}-\mathbf{R}_{0}\right|^{2}+\int_{0}^{\Delta\tau}V\left(\mathbf{R}^{L}\left(t\right)\right)\mathrm{d}t.
		\end{aligned}
		\label{linear Euclidean action}
 \end{equation}
For the cases treated here, the remaining one-dimensional integral can be performed analytically. With this,
\begin{equation}
\begin{aligned}e^{-\Delta\tau\hat{H}/\hbar}\psi\left(\mathbf{R}_{0}\right) & \simeq\left(\frac{m}{2\pi\Delta\tau\hbar}\right)^{d/2}\\
 & \times\int_{\mathbb{R}^{d}}\exp\left\{ -\frac{1}{\hbar}S_{\text{E}}^{\text{L}}\left(\mathbf{R}_{0},\mathbf{R}_{T},\Delta\tau\right)\right\} \\
 & \times\psi\left(\mathbf{R}_{T}\right)\mathrm{d}\mathbf{R}_{T}.
\end{aligned}
\label{final propagation}
\end{equation}
We use this final expression below because it results in a very simple but sufficiently effective implementation.
Nevertheless, we expect that performing a path integral with several intermediate time slices at each time step will be more efficient once some technical optimizations are carried out.

\subsection{Monte Carlo/quasi-Monte Carlo integration}
\label{mcseciton}
\paragraph*{Stochastic estimate of the path integral.}
Whether we use Eq.~\eqref{path integral}, Eq.~\eqref{new path integral}, or Eq.~\eqref{final propagation}, the evaluation entails a high-dimensional integral.
This can be carried out using an MC procedure since we will not encounter a serious sign problem as long as $\Delta \tau$ remains small.
Furthermore, unless the path passes through a negatively divergent potential, the dominant contribution to the Euclidean action arises from the kinetic term, which is an $Nd$-dimensional Gaussian function centered at $\mu=\mathbf{R}_{0}$ with a variance of $\sigma^{2}=\frac{\hbar\Delta\tau}{m}$.
This Gaussian, which we will denote as  $\mathcal{N}_{\mu,\sigma^{2}}(\mathbf{R})$, therefore has a large overlap with the integrand, making it a good but very simple weight for importance sampling.
With this in mind, the wavefunction after propagation can be evaluated numerically as follows:
\begin{equation}
\begin{aligned} & e^{-\Delta\tau\hat{H}/\hbar}\psi\left(\mathbf{R}_{0}\right)\simeq\left(\frac{m}{2\pi\Delta\tau\hbar}\right)^{Nd/2}\frac{1}{n_{s}}\\
 & ~~~~~~~~\times\sum_{\mathbf{R}_{i}\sim\mathcal{N}_{\mu,\sigma^{2}}}\frac{\exp\left\{ -\frac{1}{\hbar}S_{\text{E}}^{\text{L}}\left(\mathbf{R}_{0},\mathbf{R}_{i},\Delta\tau\right)\right\} }{\mathcal{N}_{\mu,\sigma^{2}}\left(\mathbf{R}_{i}\right)}\psi\left(\mathbf{R}_{i}\right).
\end{aligned}
\label{monte carlo}
\end{equation}
Here $\mathbf{R}_{i}$ are  $n_{s}$ random coordinates drawn from the Gaussian distribution defined by $\mathcal{N}_{\mu,\sigma^{2}}$.

\paragraph*{Quasi-Monte Carlo.}
MC integration with Gaussian importance sampling can straightforwardly be replaced by quasi-MC techniques, where instead of random coordinates $\mathbf{R}_{i}$, quasi-random ones (such as the Sobol sequence, which we use here) are used to obtain improved convergence properties \cite{press_numerical_2007}.
Here we do so by passing the uniformly distributed Sobol sequence through the quantile function of a normal distribution.

\subsection{Symmetrization}
\label{symsection}

\paragraph*{Exchange symmetry and smooth ansatzes.}
VMC necessarily requires spatially smooth ansatzes, because discontinuities in either the wavefunction or its gradient imply diverging kinetic energy.
The procedure proposed here should certainly be expected to eventually converge to an approximately smooth result, but smoothness is by no means required at intermediate stages of the algorithm.
This gives us some extra freedom in the definition of the ansatz, which turns out to be useful for enforcing symmetries.

Let us define the problem concretely for particle exchange symmetry.
Consider a set of coordinates $\mathbf{R}=\left( \mathbf{r}_{1},\ldots,\mathbf{r}_{N}\right)$ representing the locations of a set of $N$ indistinguishable particles in $d$ dimensions.
Given a permutation $\bar{\pi}$ of the coordinates, and denoting $\bar{\pi}\mathbf{R}\equiv\left(\mathbf{r}_{\bar{\pi}_{1}},\ldots,\mathbf{r}_{\bar{\pi}_{N}}\right)$, particle exchange symmetry requires that 
\begin{equation}
	\psi\left(\mathbf{R}\right)=\sigma\left(\bar{\pi}\right)\psi\left(\bar{\pi}\mathbf{R}\right).
	\label{symmetry}
\end{equation}
Here, $\sigma\left(\bar{\pi}\right)$ is unity for bosons and the sign of the permutation for fermions.

In the derivative-based SRW method of Ref.~\cite{Atanasova2023}, time propagation is performed by applying a linearized evolution operator, $e^{-\Delta\tau\hat{H}/\hbar} \simeq 1-\frac{\Delta\tau}{\hbar}\hat{H}$, to the wavefunction.
Since the Hamiltonian includes a kinetic energy term, this in turn requires applying the Laplacian operator to the wavefunction, which must therefore be smooth.
With this in mind, in Ref.~\cite{Atanasova2023} we proposed to deal with exchange symmetry by stochastic projection: given a set of sampled coordinates and wavefunction values, we generated new samples with permuted coordinates and appropriate signs.
While this works, it appears to entail an unnecessary combinatorial complexity: the wavefunction must be learned in $N!$ regions that differ from each other only by a sign, and explicitly learning the sign structure seems needlessly difficult \cite{hutter_representing_2020}, as it is supposedly a manifestation of a rather simple rule.
The previous approach relied exclusively on machine learning to figure out the entire space, even though all of the information is located in any one of the $N!$ regions.
In practice, we found it expensive to scale that approach beyond 4--5 particles in the cases presented below.

\paragraph*{Sorting-based exchange symmetry.}
When time propagation is performed by path integration, smoothness requirements can be relaxed.
A computationally efficient way to enforce particle exchange symmetry/antisymmetry is then based on simply sorting the coordinates before evaluation \cite{sannai_improved_2021,hutter_representing_2020}; see also \cite{richter-powell_sorting_2023} for a related idea.
This is not a generally usable approach within VMC or derivative-based SRW, because it results in discontinuous functions; but it is perfectly valid when using path-integral-based SRW.
Notably, a related approach is viable when using explicit sparse grid ansatzes, which are discrete by nature \cite{yserentant_sparse_2006,griebel_sparse_2007}.

To enforce a sign structure, we first define a unique ordering of the coordinates $\mathbf{r}_n$.
One possible way to order the $d$-dimensional vectors $\mathbf{r}_{i}$ is lexicographically: for example, a 3D vector $\mathbf{r}_{i}=\left\{ x_{i},y_{i},z_{i}\right\}$ is deemed smaller than $\mathbf{r}_{j}$ if $x_{i}<x_{j}$, or if $x_{i}=x_{j}$ and $y_{i}<y_{j}$, or if $x_{i}=x_{j}$ and $y_{i}=y_{j}$ and $z_{i}<z_{j}$.
Now, whenever evaluating $\psi(\mathbf{R})$,  we first sort the components of $\mathbf{R}$,  thereby finding the permutation $\bar{\pi}(\mathbf{R})$ such that $\bar{\pi}(\mathbf{R})\mathbf{R}$ is in the unique orthant corresponding to fully sorted coordinates.
We can then perform function evaluation only in that orthant by using Eq.~\eqref{symmetry} so that the permutation signs are used to exactly enforce exchange symmetry.

The computational complexity of function evaluation with sorting-based exchange symmetry enforcement is set by that of the sorting algorithm, e.g. $\mathcal{O}\left(N\log N\right)$ when using quicksort.
This should be compared to a complexity of $\mathcal{O}\left(N^{3}\right)$ for evaluating a Slater determinant and $\mathcal{O}\left(N^{2}\right)$ for the Vandermonde approach \cite{hutter_representing_2020,pang_on2_2022}.
Within the orthant we can rely on the universality of implicit neural representations of functions \cite{hornik_multilayer_1989}, so it can also be compared to the $\mathcal{O}\left(N^{2}\log{N}\right)$ cost of a linear set of ``sortlets'' \cite{richter-powell_sorting_2023}.

We also note that enforcement of exchange symmetry by sorting can easily be generalized to systems with several different types of indistinguishable particles, such as mixtures of fermions and bosons or mixed spin configurations.
This, by considering permutations of only subsets of the coordinates.
An analogous procedure can also be used to enforce spatial symmetries, and we will do so below.

\subsection{Regression}
\label{regsection}
\paragraph*{Learning the wavefunction from samples.}
At each step of the SRW calculation, the propagated wavefunction is evaluated at $M$ sets of coordinates to obtain the samples $\left\{ \mathbf{R}_{i},e^{-\Delta\tau\hat{H}/\hbar}\psi\left(\mathbf{R}_{i}\right)\right\}$.
Note that here $\psi\left(\mathbf{R}\right)$ is assumed to be a \emph{known} function: either the initial condition in the first iteration, or a function that was learned in the previous step.
Learning is performed only in the fully sorted symmetry orthant using Eq.~\eqref{symmetry}.
To enable this, all samples are replaced by their symmetrized counterparts
 $\left\{ \bar{\pi}_i\mathbf{R}_i,\sigma_{i}e^{-\Delta\tau\hat{H}/\hbar}\psi\left(\mathbf{R}_{i}\right)\right\} $,
where $\bar{\pi}_i\equiv\bar{\pi}(\mathbf{R}_i)$ is the permutation that sorts $\mathbf{R}_{i}$ and $\sigma_i \equiv \sigma(\bar{\pi}_i)$ is the sign associated with it (see subsection \ref{symsection}).
This data is used to train a model $\psi_{\theta}^{\Delta\tau}(\mathbf{R})$, where $\theta=\mathrm{argmin}_\theta~J\left(\theta\right)$ is a set of parameters to be learned by minimizing a loss function $J(\theta)$.
In the next iteration of the calculation, the wavefunction will then have the new form
\begin{equation}
\psi^{\Delta\tau}\left(\mathbf{R}\right):=\sigma\left(\bar{\pi}(\mathbf{R})\right)\psi_{\theta}^{\Delta\tau}\left(\bar{\pi}(\mathbf{R})\mathbf{R}\right).
\end{equation}
We used the superscript $\Delta\tau$ to emphasize the construction of the wavefunction $\psi^{\Delta\tau}\left(\mathbf{R}\right)$ that will be used in the next step from $\psi_{\theta}^{\Delta\tau}\left(\mathbf{R}\right)$, once it has been learned in the present step.

It is not clear whether an optimal choice for $J(\theta)$ exists, and different loss functions may result in different convergence properties and levels of accuracy.
The loss we used is based on the mean squared relative error between the model and samples:
\begin{equation}
J\left(\theta\right)=\frac{1}{M}\sum_{i=1}^{M}\frac{\left(\psi_{\theta}^{\Delta\tau}\left(\bar{\pi}_{i}\mathbf{R}_{i}\right)-\sigma_{i}e^{-\Delta\tau\hat{H}/\hbar}\psi\left(\mathbf{R}_{i}\right)\right)^{2}}{\left|e^{-\Delta\tau\hat{H}/\hbar}\psi\left(\mathbf{R}_{i}\right)\right|^{2}+\epsilon}.\label{eq:loss}
\end{equation}
The regularization parameter $\epsilon$ is added to avoid numerical instability; in all cases below, we take $\epsilon=10^{-3}$.
Taking the relative error rather than the absolute error helps in avoiding underfitting in regions where the wavefunction is small, such as the nodes.

We note that during the sample selection stage, the metropolis algorithm was used to draw $70\%$ of samples from the wavefunction's probability distribution.
The remaining samples were chosen from a uniform distribution within a hypercube large enough to encompass the region where the wavefunction is not negligible, so as to ensure that regions where it is essentially zero don't end up with random values and assist with potential ergodicity issues.
This is most likely not a crucial aspect of the algorithm.

\paragraph*{Alternative to normalization.}
Imaginary time propagation multiplies the wavefunction by exponential factors, destroying its normalization.
While the SRW method does not require the wavefunction to be normalized, eventually this can result in numerical problems.
Normalization, on the other hand, requires the evaluation of an expensive integral.
To avoid both issues, after each propagation step we scale the wavefunction so that the largest value found, after rejecting outliers, is approximately unity.

\subsection{Energy estimation}
\label{enrgsection}
We have proposed to model wavefunctions by ansatzes that may not be continuous.
This means they cannot be used to obtain a variational estimate for the energy.
While the estimate could be calculated in principle by evaluating $\left\langle \psi\right|\hat{H}\left|\psi\right\rangle /\left.\left\langle \psi\right|\psi\right\rangle $ by MC integration, it would generally be infinite.
We therefore propose two other ways to estimate the energy.
The first of these is mainly used to determine whether the ansatz has converged to the ground state, and is not variational; while the second provides a variational bound based on a smooth approximation to the non-smooth ansatz.
Both techniques avoid the need to explicitly take derivatives of the wavefunction.

\paragraph*{Phase/decay tracking.}
The first estimator we will present converges to the ground state energy when the ground state wavefunction is acquired.
To motivate its introduction, consider the following property of the exact ground state $\varphi_0$.
When propagated by an imaginary (or real) time, it is modified only by an exponential decay (or phase) factor.
Concentrating on the imaginary case,
\begin{equation}
e^{-\Delta\tau\hat{H}/\hbar}\varphi_{0}\left(\mathbf{R}\right)=e^{-\Delta\tau E_{0}/\hbar}\varphi_{0}\left(\mathbf{R}\right).
\label{ground state prop}
\end{equation}
This can be rewritten as
\begin{equation}
E_{0}=-\frac{\hbar}{\Delta\tau}\ln\left(\frac{e^{-\Delta\tau\hat{H}/\hbar}\varphi_{0}\left(\mathbf{R}\right)}{\varphi_{0}\left(\mathbf{R}\right)}\right).
\label{ground state functional}
\end{equation}
The latter is correct for any choice of coordinate $\mathbf{R}$; it therefore remains true when averaged over coordinates drawn from some distribution $P\left(\mathbf{R}\right)$:
\begin{equation}
	E_{0}=-\frac{\hbar}{\Delta\tau}\left\langle \ln\left(\frac{e^{-\Delta\tau\hat{H}/\hbar}\varphi_{0}\left(\mathbf{R}\right)}{\varphi_{0}\left(\mathbf{R}\right)}\right)\right\rangle _{P\left(\mathbf{R}\right)}.
	\label{ground state average}
\end{equation}

With this in mind, we define the following quantity:
\begin{equation}
E_\text{decay} \left[\psi\right]=-\frac{\hbar}{\Delta\tau}\left\langle \ln\left(\frac{e^{-\Delta\tau\hat{H}/\hbar}\psi\left(\mathbf{R}\right)}{\psi\left(\mathbf{R}\right)}\right)\right\rangle _{P\left(\mathbf{R}\right)}.
\label{functional}
\end{equation}
For the exact ground state, the result will be independent of the distribution $P(\mathbf{R})$.
Generally, we can expect better approximations to the ground state wavefunctions to produce more accurate energies; but these energies need not obey a variational principle, and therefore do not provide an upper bound for the ground state energy.
The evaluation of this Eq.~\eqref{functional} can typically be done with a relatively small number of samples compared to that needed for regression or variational energy estimation.
It therefore does not incur a meaningful computational cost.
Furthermore, each sampled coordinate $\mathbf{R}_i$ provides an independent estimator for the energy; a distribution of all estimators can therefore be obtained.
This enables two separate tests of convergence with respect to the imaginary time: when approaching the ground state, the mean of the estimator distribution should become constant, and its variance should approach zero.
The error plotted for each energy is taken to be the standard error of the average energies of the samples used.

We chose to use the same $M$ sets of coordinates previously used to perform the regression for this step.
Notably, very small values of the wavefunction result in large stochastic errors and the largest values are also prone to numerical instabilities.
We therefore found that higher accuracy can be attained by (arbitrarily) including only coordinates that when ordered by absolute wavefunction value, lie between the $40^\text{th}$ and $90^\text{th}$ percentiles.

\paragraph*{Restoring the variational property.}
To obtain a variational estimate, we require a smooth function approximating our non-smooth ansatz.
A straightforward way to obtain one is by convolving the ansatz with a smooth kernel, such as a Gaussian:
\begin{equation}
\begin{aligned}\psi\left(\mathbf{R}\right)\rightarrow\widetilde{\psi}\left(\mathbf{R}\right) & =\psi\left(\mathbf{R}\right)*\mathcal{N}_{0,\sigma^{2}}\left(\mathbf{R}\right)\\
	& =\int_{\mathbb{R}^{d}}\psi\left(\mathbf{R}-\mathbf{R}^{\prime}\right)\mathcal{N}_{0,\sigma^{2}}\left(\mathbf{R}^{\prime}\right)d\mathbf{R}^{\prime}.
\end{aligned}
\label{convo}
\end{equation}
The energy of the smoothed wavefunction $\widetilde{\psi}\left(\mathbf{R}\right)$ is now well-defined for any finite value of $\sigma$.
Utilizing the properties of the derivative of a convolution, this can be done without taking derivatives of $\psi\left(\mathbf{R}\right)$.
In particular,
\begin{equation}
\begin{aligned}
\nabla^{2}\left[\psi\left(\mathbf{R}\right)*\mathcal{N}_{0,\sigma^{2}}\left(\mathbf{R}\right)\right]	&=\left[\psi\left(\mathbf{R}\right)*\nabla^{2}\mathcal{N}_{0,\sigma^{2}}\left(\mathbf{R}\right)\right]
	\\&=\left[\psi\left(\mathbf{R}\right)*\left(\frac{\mathbf{R}^{2}-\sigma^{2}}{\sigma^{4}}\right)\mathcal{N}_{0,\sigma^{2}}\left(\mathbf{R}\right)\right].
\end{aligned}
\label{der_of_conv}
\end{equation}
Inserting this into the variational estimate for the energy of  $\widetilde{\psi}\left(\mathbf{R}\right)$ gives an expression that can be evaluated by MC integration:
\begin{equation}
E_{\text{MC}}=\frac{\left\langle \widetilde{\psi}\left|\hat{H}\right|\widetilde{\psi}\right\rangle }{\left\langle \widetilde{\psi}\right.\left|\widetilde{\psi}\right\rangle }\simeq\frac{\sum_{\mathbf{R},\mathbf{R}^{\prime},\mathbf{R}^{\prime\prime}}A\left(\mathbf{R},\mathbf{R}^{\prime},\mathbf{R}^{\prime\prime}\right)}{\sum_{\mathbf{R},\mathbf{R}^{\prime},\mathbf{R}^{\prime\prime}}B\left(\mathbf{R},\mathbf{R}^{\prime},\mathbf{R}^{\prime\prime}\right)},
\label{convo_energy}
 \end{equation}
where
\begin{equation}
\begin{aligned}A\left(\mathbf{R},\mathbf{R}^{\prime},\mathbf{R}^{\prime\prime}\right) & \equiv B\left(\mathbf{R},\mathbf{R}^{\prime},\mathbf{R}^{\prime\prime}\right)\\
	& \times\frac{\mathbf{R}^{\prime\prime2}-\sigma^{2}+\sigma^{4}V\left(\mathbf{R}\right)}{\sigma^{4}},\\
	B\left(\mathbf{R},\mathbf{R}^{\prime},\mathbf{R}^{\prime\prime}\right) & \equiv\frac{\psi^{*}\left(\mathbf{R}-\mathbf{R}^{\prime}\right)\psi\left(\mathbf{R}-\mathbf{R}^{\prime\prime}\right)}{\left|\psi\left(\mathbf{R}\right)\right|^{2}}.
\end{aligned}
\label{A and B}
\end{equation}
Here $\mathbf{R}$ is drawn from the wavefunction's distribution using the Metropolis algorithm,
while $\mathbf{R}^{\prime},\mathbf{R}^{\prime\prime}\sim\mathcal{N}_{0,\sigma^{2}}$
are drawn from a normal distribution with mean zero and standard deviation
$\sigma$. The error of this calculation is similar to that in a standard Monte Carlo integration; it is the standard error derived from multiple realizations of the integral.

The downside of this process is that we must perform integration over a higher-dimensional space, $\mathbb{R}^{3Nd}$; and that the original ansatz $\psi(\mathbf{R})$ is ``smeared'' across an arbitrary length scale $\sigma$ that controls the smoothness of $\widetilde{\psi}$.
The advantage is the restoration of the variational property, in that necessarily $E_\mathrm{MC} \ge E_0$.
In principle, $\sigma$ can be taken to be a variational parameter; but we can also simply choose a value that is smaller than other length scales in the problem; but large enough to avoid large contributions to the kinetic energy due to remnants of discontinuity in $\psi(\mathbf{R})$. We found a value of $\sigma=0.2$ stable and satisfactory for our purpose.

The variational energy estimator requires a substantially higher computational cost than the decay tracking strategy.
Nevertheless, it is useful in the final stage of the method, because it provides a rigorous upper bound for the energy.
Generally, the technique may be useful in a wider context: it could be applied wherever one wishes to obtain a variational estimate from a non-differentiable wavefunction.

\subsection{Implementation details\label{subsec:implementaion_details}}
As a model architecture, we chose a straightforward dense NN with 6 layers, each with 512 nodes and ReLU as the activation function.
We found ReLU-based NNs, which have a noncontinuous derivative and no second derivative, to be highly expressive and optimizable, surpassing results with smooth activation functions like ELU and Tanh.
Indeed ReLU activations are widely used in various machine learning tasks, but the need for differentiability has made them problematic to employ in VMC calculations.
In path-integral-based SRW, differentiability is not required, and ReLU activations can be safely used.
A Gaussian layer was used to enforce the boundary conditions at large distances, and in the 3D calculations an explicit Jastrow factor was applied to the ansatz.

Our implementation used Google’s TensorFlow library \cite{tensorflow2015-whitepaper}, combined with the ArrayFire library \cite{Yalamanchili2015}.
All regression tasks used the Adam optimizer \cite{kingma_adam_2017} with a geometric learning rate schedule.
Calculations were carried out on Nvidia RTX A6000 graphics processing units.

Except where specified,  all the results presented below were obtained by propagating a randomly initialized NN until the decay tracking energy estimator reached convergence within the MC errors.
Path integration is performed with $n_{s}=\num{2500}$ paths per sample, and regression with $M=\num{20000}$ samples.

\section{Results}
We will consider $N$ particles in a harmonic trap.
Our particles can be either bosons or fermions, and we will consider both 2D and 3D traps.
The fermionic version of this model, sometimes called a ``Hooke's atom'' \cite{kestner_study_1962}, has long been used to describe quasiparticles in semiconducting quantum dots \cite{reimann_electronic_2002}.
More recently, such traps have been realized in ultracold atomic systems \cite{yin_harmonically_2014,schillaci_energy_2015}, where particles can be either fermionic or bosonic.

The model is described by the Hamiltonian
\begin{equation}
	\hat{H}=-\frac{\hbar^{2}}{2m}\nabla^{2}+\frac{1}{2}m\omega^{2}\sum_{i=1}^{N}\hat{\mathbf{r}}_{i}^{2}+\sum_{i=1}^{N}\sum_{j>i}^{N}\frac{\lambda}{\left|\hat{\mathbf{r}}_{i}-\hat{\mathbf{r}}_{j}\right|}.
	\label{harmonic with lambda}
\end{equation}
Here $m$ is the mass of the particles, $\omega$ parametrizes the trap, and $\lambda$ sets the interaction strength.
We work in atomic units by setting $\hbar=m=\omega=1$; distances are therefore given in units of $\sqrt{\frac{\hbar}{m\omega}}$. 
Time propagation was performed in steps of size $\Delta\tau=0.05$ in these units, except where we state otherwise.
For both the harmonic potential and the Coulomb interaction, the integral of Eq.~\eqref{linear Euclidean action} can be solved analytically.
All our results are obtained using the architecture described in Sec.~\ref{subsec:implementaion_details}.

\subsection{Benchmark: noninteracting particles}
  \begin{figure}
	\centering
	\includegraphics[width=0.5\textwidth]{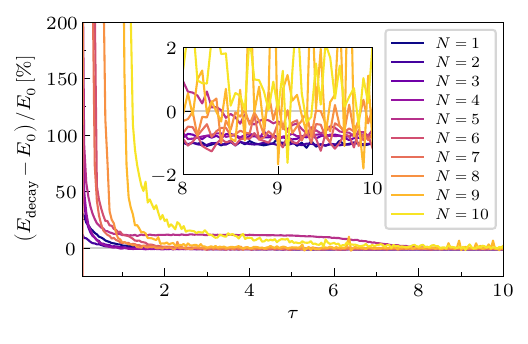}
	\caption{
		Convergence of the relative error of the decay estimator to the ground state energy with imaginary time, for $N$ spin-$\frac{1}{2}$ noninteracting fermions in the low-spin state of a 2D harmonic trap.
	}
	\label{fig:full propagation}
\end{figure}
\begin{figure}
	\centering
	\includegraphics[width=0.5\textwidth]{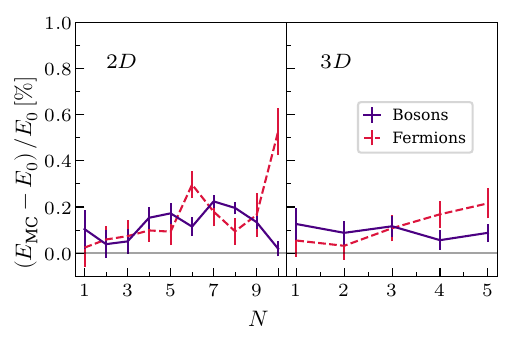}
	\caption{
		The relative error of the variational estimate of the energy at $\tau=10$, for noninteracting bosons/fermions (purple/red curves) in 2D/3D (left/right panels) harmonic wells.
		In 2D, the low-spin states of the fermions are considered; while in 3D, the high-spin states are considered.
	}
	\label{fig:errors}
\end{figure}

Since we take no explicit advantage of any specific properties of noninteracting wavefunctions, the noninteracting limit $\lambda=0$ is a useful test case.
For fermions, it also tests the effectiveness of the anti-symmetrization scheme.
Fig.~\ref{fig:full propagation} therefore shows the dependence of the decay tracking estimate for the energy, Eq.~\eqref{functional},  on imaginary time $\tau$, for up to 10 Fermionic particles in 2D and in the low-spin state.
Convergence to the exact result occurs in all cases and is asymptotically exponential until the noise threshold is reached. One exception is the case with 5 fermions, where there appears to be initial convergence followed by an eventual drop to the correct estimation.
We note that analogous behavior occurs for 3 fermions in the high spin 3D case (data not shown).
This behavior is likely influenced by the exact energy level structure and the band gap of the specific system, as well as perhaps by the energy landscape of the neural ansatz \cite{bukov_learning_2021}.
Ways to address this issue in systems where it poses a problem remain to be investigated, but a simple approach could be to verify convergence with a few uses of a variational energy estimator.
This last result appears similar to energy convergence in a VMC calculation using stochastic reconfiguration, but should not be misunderstood to have the same meaning.
In particular, decay tracking does not provide a physical expectation value of the energy unless the wavefunction comprises a single eigenstate.
Nevertheless, it is evidently not only useful for determining whether convergence has occurred, but also provides a reasonably accurate estimate for the ground state value itself. Here we observe that the converged values result in an energy consistently smaller than the exact energy. This can be explained by an undershooting of the decay in Eq. \eqref{functional}, resulting in a larger value than expected and thus a smaller energy than the exact one. This is likely an artifact of the methodology of averaging that was discussed in subsection \ref{enrgsection}. However, the MC calculations shown in Fig.~\ref{fig:errors} adhere to the variational principle. This indicates that convergence is more critical than the actual converged value when using the decay estimator.

Using our rather naive implementation and neural model, the total number of degrees of freedom $Nd$ cannot greatly exceed $\sim20$ on the graphics card we used.
The main bottleneck is graphics memory: our requirements increase linearly with $Nd$, which is necessary; but as implemented, also with the number of samples and paths.
They therefore rapidly exceed what is available on the card.
There are several technical routes to avoiding this problem without significant computational overhead, but they are beyond the scope of the present study.
Similarly, there are multiple routes to parallelizing the process over multiple cards/machines, but these will not be explored here.

\begin{table}[h!]
\centering
\begin{tabular}{|c|cc|cc|cc|cc|}
\hline
\textbf{} & \multicolumn{2}{c|}{Bosons (2D)} & \multicolumn{2}{c|}{Fermions (2D)} & \multicolumn{2}{c|}{Bosons (3D)} & \multicolumn{2}{c|}{Fermions (3D)} \\
\hline
 $N$ & $E_{\text{MC}}$ & $E_{0}$ & $E_{\text{MC}}$  & $E_{0}$ & $E_{\text{MC}}$  & $E_{0}$ & $E_{\text{MC}}$  & $E_{0}$ \\
\hline
1  & 1.0010(8) & 1 & 1.0002(8) & 1  & 1.501(1) & 1.5 & 1.500(1) & 1.5 \\
2  & 2.000(1) & 2 & 2.001(1) & 2 & 3.002(1) & 3 & 4.001(2) & 4\\
3  & 3.001(1) & 3 & 4.002(2) & 4 & 4.505(2) & 4.5 & 6.507(3) & 6.5\\
4  & 4.006(1) & 4 & 6.005(2) & 6 & 6.003(2) & 6 & 9.015(5) & 9\\
5  & 5.008(2) & 5 & 8.007(4) & 8 & 7.506(2) & 7.5 & 12.526(8) & 12.5\\
6  & 6.006(2) & 6 & 10.029(5) & 10 &   &   &   &   \\
7  & 7.015(1) & 7 & 13.023(8) & 13 &   &   &   &   \\
8  & 8.015(2) & 8 & 16.015(9) & 16 &   &   &   &   \\
9  & 9.012(2) & 9 & 19.03(1) & 19 &   &   &   &   \\
10 & 10.001(3) & 10 & 22.11(2) & 22 &   &   &   &   \\
\hline
\end{tabular}
\caption{The variational energy estimations of noninteracting bosons and fermions in 2D and 3D harmonic wells, compared to the exact energies. In 2D, the low-spin states of the fermions are considered; while in 3D, the high-spin states are considered. All values are in atomic units.}
\label{tab:particle_comparison}
\end{table}

When the decay tracking estimate converges, we evaluate a variational estimate for the energy.
The results of this are shown in Fig.~\ref{fig:errors} and Table~\ref{tab:particle_comparison}.
Here, we considered both 2D and 3D traps (left and right panels, respectively); and both $N$ spinless bosons and $N$ spin-$\frac{1}{2}$ fermions (red and purple curves, respectively).
In 2D we placed the Fermions in the low-spin state: a wavefunction with mixed spatial antisymmetry, where exchanging two particles of the same spin reverses the sign of the wavefunction, but exchanging two particles of opposite spin may not.
In 3D we targeted the high-spin state, where the spatial component of the wavefunction is fully antisymmetric in the exchange of any two particles.
Due to the model's memory scaling, as mentioned above, in 3D we treated up to 5 particles.

The variational energy is more precise, but also more computationally expensive to obtain than the decay tracking estimate; and provides a rigorous upper bound for the ground state energy to within the MC errors.
In all combinations of dimension and exchange statistics, the relative error is largely independent of the number of degrees of freedom.

\subsection{Benchmark: strong interactions}
The noninteracting system is analytically solvable.
While the agreement of our calculations with theory in the noninteracting limit are encouraging, one could suspect that some hidden aspect of the method makes that case easier within our formalism.
It is therefore informative to validate the method in another limit, where interactions are relatively strong; but not so strong as to wipe out all quantum effects.
Here, no analytical solution is known, but a variety of results from established numerical methods exists.

Ref.~\cite{chin_solving_2020} recently presented results for same-spin fermions in a 2D quantum dot at $\lambda=8$. 
They performed comparisons between a method they proposed and a variety of state-of-the-art techniques:  configuration interaction (CI) \cite{rontani_full_2006}, PIMC \cite{B._Reusch_2003},  DMC \cite{pederiva_diffusion_2000,ghosal_incipient_2007} and a ground state path integral algorithm with a  $4^\text{th}$ order propagator (GSPI4) \cite{chin_high-order_2015}.
We will present a similar comparison, with respect to our variational energy estimator.
To obtain the results used for this comparison, the imaginary time needed to obtain convergence ranged from $\tau=4$ to $\tau=8$.
We also used $\Delta\tau=0.01$ and $n_{s}=5000$.

\begin{figure}
	\centering
	\includegraphics[width=0.5\textwidth]{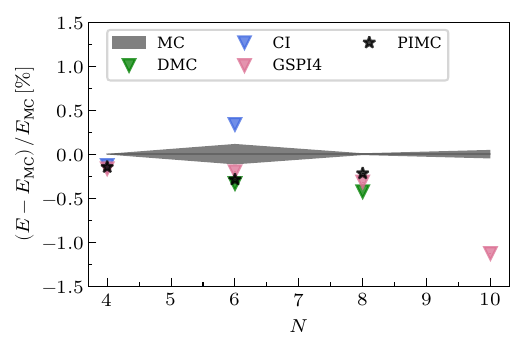}
	\caption{
		The relative difference between the converged variational estimate of the energy and the energy as calculated with different methods, for $N$ interacting high-spin fermions with $\lambda=8$.
		Several methods (different colors and symbols; see text) are compared, with Eq.~\eqref{convo_energy} defining the baseline, and the size of the shaded gray region defining the MC uncertainty for this baseline.
		Bottom-facing triangles of all colors represent upper bounds for the ground state energy.
	}
	\label{fig:exact}
\end{figure}

Our findings, shown in Fig.~\ref{fig:exact}, are consistent with the known results: differences are within $\sim 0.5\%$ for the smaller systems and $\sim 1.25\%$ for the 10 particle system.
In all cases, our result produces a higher energy than the lowest variational estimate (triangular symbols of all colors).
We do not exceed existing methods in precision within this regime, but note that (a) the model and implementation remain very naive, and (b) both accuracy and precision could be improved somewhat by using our solution as the starting point for fixed node DMC.
In fact, the VMC data underlying the results from Ref.~\cite{pederiva_diffusion_2000} provides energies of a quality similar to ours, despite its use of a specialized ansatz containing a great deal more physical insight than the neural network employed here.

\subsection{Computational scaling}
\begin{figure}
	\centering
	\includegraphics[width=0.5\textwidth]{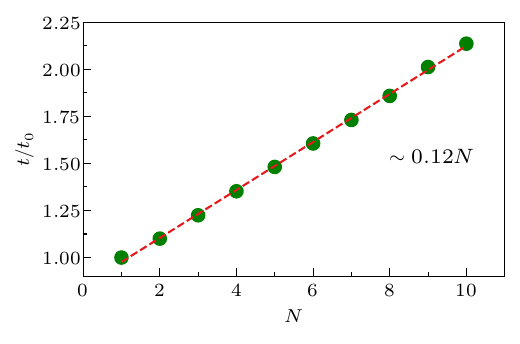}
	\caption{
		Total runtime for propagating $N$ noninteracting fermions in a 2D harmonic trap to $\tau=10$ on a single graphics card, in units of $t_{0} \equiv \num{5561}\text{ seconds}$.
		The parameters are as in Fig.~\ref{fig:full propagation}.
		}
	\label{fig:time}
\end{figure}

\paragraph*{Computational scaling with particle number.}
We now discuss some of the more universal aspects of the cost of performing simulations with the method proposed here, while attempting to avoid implementation-dependent issues.
Initially, we investigate how the total runtime to propagate a wavefunction to $\tau=10$ depends on the number of particles, $N$; this is plotted for  $N\le10$ noninteracting fermions in 2D in Fig.~\ref{fig:time}.
The result is consistent with linear scaling.

This is reasonable, as we have constructed the method in such a way that the only task whose cost increases more than linearly in $N$ is the symmetrization procedure's lexicographic ordering.
The sorting algorithm's cost scales as $\mathcal{O}\left(N\log N\right)$ and this operation's runtime should eventually become dominant; however, its prefactor is small enough that it has no discernible impact for $N\le10$.
Furthermore, the linear scaling will certainly be broken by the limited expressiveness of the model and by increasing MC errors at larger values of $N$, and gapless systems will typically require larger imaginary times as the system grows.
All this is beyond our present implementation, which is limited by graphics memory to $\sim10$ particles in 2D, as noted above.

While our longest runtimes are several hours on a single A6000 card, parallelization on recent graphics card servers with, say, 8 Nvidia H100 cards should cut this down to minutes.
Even if the memory limitations of our preliminary implementation, and assuming each card must contain all the data rather than it being distributed, this would also allow approximately quadrupling the number of degrees of freedom (i.e. $\sim40$ particles in 2D); a distributed algorithm would then enable a factor of $\sim 30$ (i.e. $\sim 300$ particles in 2D) on a single machine.

This analysis is encouraging, but most likely highly optimistic: it assumes that no new challenges will appear at larger dimensionalities.
As we discuss below, examining the fidelity of noninteracting wavefunctions suggests that this assumption should, at the very least, be taken with a grain of salt.

\paragraph*{Comparison to other methods.}
It is useful to place the method proposed here within the context of existing frameworks for treating quantum systems in real space, in order to understand where it could be useful.
The method with the most overlap to SRW is VMC using the stochastic reconfiguration for optimization, as both try to find the ground state wavefunction of the system in terms of an ansatz representation.
The two methods end up having similar memory costs if matrix-free algorithms are used to perform the stochastic reconfiguration \cite{vicentini_netket_2022}.

The computational cost of the stochastic reconfiguration step in VMC scales like $\mathscr{O}\left(P^{3}\right)$ with respect to the number of parameters $P$, precluding the use of models with a large number of parameters.
A very recent alternative instead precludes many samples \cite{chenEmpoweringDeepNeural2024}: it scales like $\mathcal{O}\left(P^{2}M+M^{3}\right)$ where $M$ is the number of samples.
In contrast, the SRW can use modern linear-scaling optimizers like Adam, and for typical deep learning architectures the cost scales like $\mathcal{O}\left(PM\right)$.

An additional computational cost is with respect to the number of particles $N$.
Generally first-quantized methods for fermionic systems use a basis of Slater determinants, which introduces a factor of $\mathscr{O}\left(N^{3}\right)$ to the scaling.
An alternative with some tradeoffs is Vandermonde approaches \cite{pang_on2_2022}.
In the sorting-based approach used here, this cost scales like $\mathscr{O}\left(N\log N\right)$.

\begin{figure}
	\centering
	\includegraphics[width=0.5\textwidth]{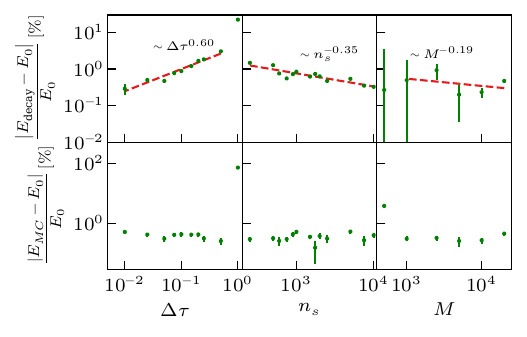}
	\caption{
		The relative error of the decay tracking energy estimator (top panels) and the variational estimate (bottom panels) for 4 noninteracting fermions in a 2D harmonic trap in the high-spin configuration, propagated to $\tau=5$.
		Columns show the dependence on different numerical parameters: the time step $\Delta\tau$ (left panels), the number of paths $n_s$ in the time propagation (middle panels) and the number of samples $M$ in the regression (right panels).
		The dashed red lines are power-law fits performed after the removal of outliers.
		Other parameters are as in Fig.~\ref{fig:full propagation}.
	}
	\label{fig:scaling}
\end{figure}

\paragraph*{Numerical convergence.}
Next, we explore the robustness of the method to its numerical parameters.
To illustrate the various dependencies, we consider 4 noninteracting fermions in a 2D harmonic trap in the high-spin (fully antisymmetric) configuration, propagated to $\tau=5$.
We modify one parameter from its previous value at a time, with the others held constant at the values used in Fig.~\ref{fig:full propagation}.
Fig.~\ref{fig:scaling} shows how our two energy estimators (top and bottom panels) depend on the imaginary time step $\Delta\tau$ (left panels), the number of linear paths $n_{s}$ in the path integration (middle panels) and the number of samples $M$ used in the regression steps (right panels).
A fitted power law is also shown in each top panel.

The decay tracking estimate (top panels in Fig.~\ref{fig:scaling}) is sensitive to all three numerical parameters.
$\Delta\tau$ appears to have the strongest impact: the error grows with close to a square root power law.
The error with respect to $M$ and $n_s$ decreases with smaller power laws.
In both cases, this performance is inferior compared to what one would expect from a single MC or quasi-MC calculation.
This is not surprising, since errors are propagated through the calculation in time; but may not be the asymptotic limit at large sample sizes.

Changing the number of regression samples $M$ has a larger effect on the error bar of the decay tracking estimate than on its average value.
The relatively precise, if noisy, energy values at low sample numbers suggest that the imaginary time propagation progressively casts out noise and artifacts that are generated by noisy regression, providing some additional stabilization to the process.

Interestingly, the numerical parameters have little to no effect on the variational energy (bottom panels in Fig.~\ref{fig:scaling}), except in that very large values of $\Delta\tau$ and very small sample numbers $M$ result in convergence to a wrong result.
Remarkably, even a very small number of paths is sufficient to provide reasonable accuracy.
We also note that since in the variational estimate, the wavefunction is effectively convolved with a Gaussian of finite width, it can sometimes produce energies that are less accurate than those generated by the decay tracking estimator.
Nevertheless, this estimator has the important advantage of providing an upper bound on the ground state energy.

\subsection{Wavefunction fidelity}
\begin{figure}
	\centering
	\includegraphics[width=0.5\textwidth]{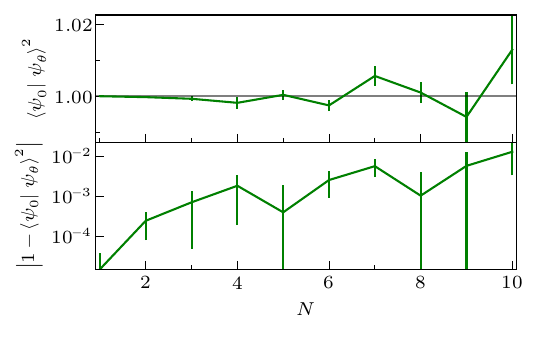}
	\caption{
		(Top) The fidelity of our ansatz when fitted to the exact wavefunction of $N$ noninteracting fermions in the low-spin ground state.
		(Bottom) The corresponding infidelity on a logarithmic scale.}
	\label{fig:fidelity}
\end{figure}
We have shown that the rather simplistic neural model we have chosen is expressive enough to produce reasonably accurate energies at high dimensionalities.
This is not entirely surprising.
Real space VMC methods based on neural networks have already shown better accuracy than ours for larger systems: for example, Ref.~\cite{schatzle_deepqmc_2023} presents hydrogen chains with up to 45 electrons in 3D as benchmarks).
This, despite a higher theoretical computational scaling $\sim N^4$ in that method, which is not yet encountered for tractable system sizes, which scale better in practice.
However, these methods are based on more refined architectures that incorporate physical and chemical knowledge into the ansatz.
It therefore seems prudent to seek out a more sensitive probe of the eventually inevitable breakdown of the model.

Perhaps the most formidable test for a high-dimensional ansatz is fidelity with respect to the wavefunction it aims to approximate.
The breakdown we are attempting to observe is a property of the wavefunction and model alone, and has nothing to do with the time propagation.
We therefore examined the fidelity $\left|\left.\left\langle \psi_{0}\right|\psi_{\theta}\right\rangle \right|^{2}$  between the exact noninteracting fermionic ground state of the low-spin 2D system, $\psi_0$; and our model $\psi_{\theta}$ after training on samples distributed similarly to the procedure we used in practice.
We do not optimize the wavefunction explicitly for fidelity, which is difficult with an ansatz that isn't intrinsically normalized.
As in the algorithm described above, we minimize the loss in Eq.~\eqref{eq:loss}.

For each value of $N$, we performed 50 independent regression procedures for 60 independent sets of samples and evaluated the fidelity for each regression by MC integration.
Since our ansatz may not be perfectly normalized, the result for any particular regression can be slightly larger than 1.
The top panel of Fig.~\ref{fig:fidelity} shows the average over regressions of the fidelity as a function of $N$, with error bars given by the standard error.
It is clear that with more particles, the distribution widens somewhat, making it less likely that a high fidelity was obtained.

To allow for a more quantitative analysis of the difficulty of reaching large particle numbers with the ansatz and fitting procedure used here, it is useful to plot the infidelity on a logarithmic scale (see bottom panel of Fig.~\ref{fig:fidelity}).
The latter is defined here as the absolute value of the difference between the fidelity and 1.
While increasing exponentially at small particle numbers, the rise in the level of infidelity as $N$ increases eventually appears to taper off at values $\sim\num{0.01}$.
This apparently sub-exponential growth is promising: it suggests that the method may well be scalable to substantially larger system sizes.
This result does not take from the fact that we tested a very limited neural network architecture. A dense neural network is perhaps the simplest design possible for this use case. Various new studies demonstrate architectures specifically designed for wavefunctions of this kind \cite{lange_architectures_2024}, integrating them with the suggested methodology will surely improve fidelity and other measurable quantities. The ease of switching architectures mid-process highlights the endless possibilities for such combinations.

\subsection{Weakly interacting limit}
\begin{figure}
	\centering
	\includegraphics[width=0.5\textwidth]{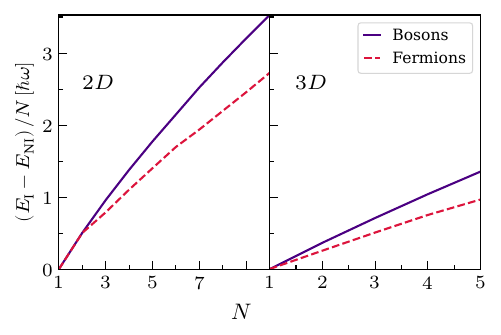}
	\caption{
		The converged interaction energy per particle for a harmonic trap, with particles interacting at strength $\lambda=1$.
		Solid lines denote spinless bosons, while dashed lines denote spin-$\frac{1}{2}$ fermions.
		In 2D (left panel), the low-spin state of the fermions is considered, while in in 3D (right panel), the high-spin state is considered.
		The error bars are too small to see.
	}
	\label{fig:energy diff}
\end{figure}
Having established a general picture of the method's reliability, we now set $\lambda=1$ and explore several examples showing the effect of interactions on the many-body system.
Fig.~\ref{fig:energy diff} shows the difference between the converged ground state energy per particle and its noninteracting counterpart.
This is plotted at several particle numbers $N$, for a 2D (left panel) and 3D (right panel) quantum dot; and for spinless bosons (solid line) and spin-$\frac{1}{2}$ fermions (dashed line).
To demonstrate the flexibility of the method, we consider the low-spin fermionic state obeying Hund's law in 2D; whereas in 3D, we examine the high-spin state, where the spatial part of the wavefunction is fully antisymmetric with respect to particle exchange.
Imaginary time propagation was performed up to $\tau=10$.
 
 The interaction energy per particle increases almost linearly with the number of particles and is stronger for 2D containment in all cases, as might be expected.
 Generally, bosons gain more interaction energy than fermions due to the lack of Pauli repulsion (disregarding the $N=1$ case and the low-spin case with $N=2$, where bosons and fermions cannot be distinguished from each other).
 The dependence on particle number is otherwise rather featureless at this interaction strength.

\begin{figure}
    \centering
    \includegraphics[width=0.5\textwidth]{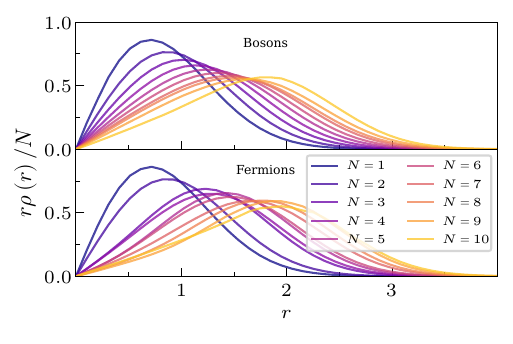}
    \caption{
    	Radial density per particle of spinless bosons (top) and low-spin fermions (bottom) in a 2D harmonic trap, at interaction strength $\lambda=1$.
    	Parameters correspond to those in the left panel of Fig.~\ref{fig:energy diff}.
    }
    \label{fig:interacting_F_and_B}
\end{figure}

Next, in Fig.~\ref{fig:interacting_F_and_B}, we present the angle-integrated radial density per particle $r\rho(r)/N$ at the same parameters used to produce the left panel of Fig.~\ref{fig:energy diff}.
Here $\rho\left(r\right)\equiv\int\mathrm{d}\phi n\left(r,\phi\right)$, where
\begin{equation}
	n\left(r,\phi\right)=n\left(\mathbf{r}\right)\equiv\int\mathrm{d}\mathbf{r}_{2}\cdots\mathrm{d}\mathbf{r}_{N}\left|\psi\left(\mathbf{r},\mathbf{r}_{2},\ldots,\mathbf{r}_{N}\right)\right|^{2}.
\label{eq:density}
\end{equation}
The area under each curve is therefore unity.
For the fermionic cases in which spins are unequally occupied, so that density is spin-dependent, we average over spin.
As the number of particles $N$ increases, the bosonic system (top panel of Fig.~\ref{fig:interacting_F_and_B}) progressively and smoothly expands outwards.
The fermionic system (bottom panel of the same figure) is in a low-spin state, and the spin-degeneracy of the noninteracting orbitals expresses itself in the existence of several pairs of radial densities with similar or identical asymptotic behavior near the origin.

In most cases, the interacting solutions inherit an infinite rotational degeneracy of the ground state from their noninteracting counterparts, and are not uniquely defined.
Of particular interest are the cases with closed shells, where the noninteracting ground state is unique and characterized by a density with no dependence on angle.
Any angular dependence in the interacting solution is then due to spontaneous symmetry breaking caused by the many-body interactions.
We will consider such scenarios next.

\subsection{Strong interactions and symmetry breaking}
\begin{figure*}
  \includegraphics[width=\textwidth]{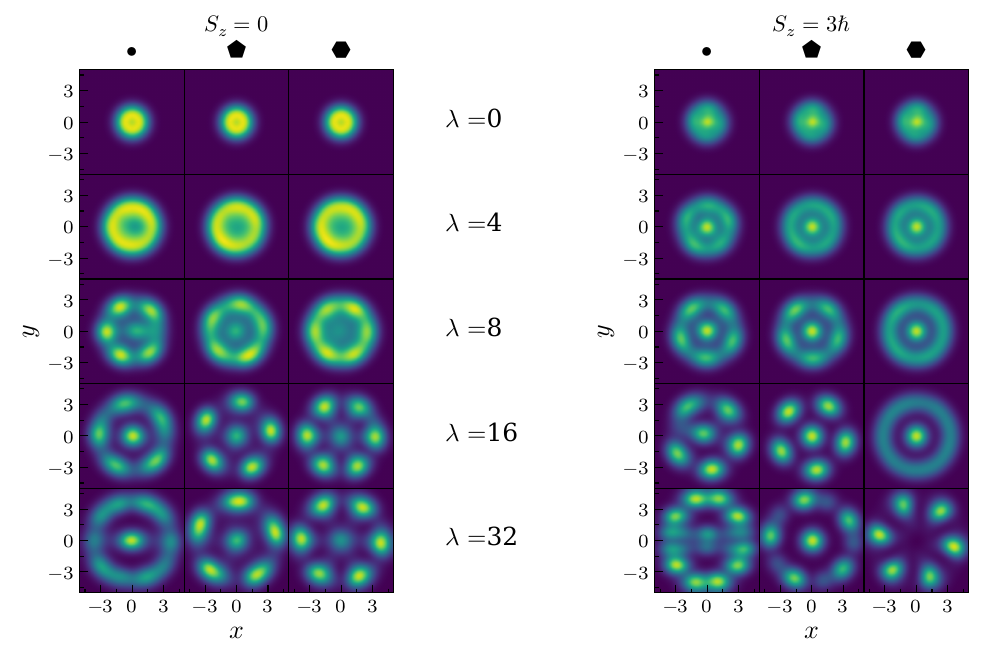}
  \caption{
  	The density of 6 fermions in a 2D harmonic trap, for the low-spin (left panel) and high-spin (right panel) configurations.
  	The interaction strength $\lambda$ is varied in rows, increasing from top to bottom.
  	Either no specific symmetry was enforced (left columns of subpanels),  5-fold rotational symmetry was enforced (middle columns) or 6-fold rotational symmetry was enforced.
  	}
\label{fig:wigner}
\end{figure*}

Finally, we will discuss the application of our approach to the ground state of six fermions in a 2D trap.
In Fig.~\ref{fig:wigner} we show the effect of varying the interaction strength $\lambda$, increasing from top to bottom, on the spin-averaged density (see Eq.~\eqref{eq:density}) in both the low-spin (left panel) and high-spin (right panel) cases.
As we previously mentioned, in the absence of interactions ($\lambda=0$, top subpanels) the ground state of this system forms a closed shell, and therefore exhibits radial symmetry and no degeneracy.
However, interactions may spontaneously break this symmetry, thus generating an infinitely degenerate interacting ground state and potentially an even larger manifold of nearby excited states.
In the left subpanels of each panel in Fig.~\ref{fig:wigner} we show that symmetry breaking indeed happens for both spin configurations.
However, the densities corresponding to the approximate ground state that was found by our algorithm are an arbitrary superposition of states from the ground state manifold, as well as other states with indistinguishably small excitation energies.
The results therefore feature no particular symmetry and are difficult to interpret.
Furthermore, they are hard to reconcile with intuition from the semiclassical limit at large interaction strength, where Wigner molecules are eventually expected to form.

Nevertheless, one can observe that both spin configurations assume a shell-like structure at weak interaction strength, with the polarized case exhibiting a central bump that steps from the noninteracting orbital structure.
Around $\lambda=8$, an angular dependence begins to develop and the differences between the two spin configurations become less pronounced.
Similar behavior has been predicted in previous theoretical investigations of the formation of Wigner molecules in 2D quantum dots \cite{egger_crossover_1999,ghosal_incipient_2007,xie_ab-initio_2022}.

In order to obtain more comprehensible densities that can be easily interpreted, we can impose an additional spatial symmetry of our choice on the system, using a method similar to the lexicographic symmetrization of section~\ref{symsection}.
For a given symmetry operation of order $n$ that commutes with the Hamiltonian, we partition the 2D physical space into $n$ regions comprising an orbit, and number them.
For a rotation the regions are sextants.
We can then assign to any configuration $\left\{ \mathbf{r}_{1},\ldots,\mathbf{r}_{N}\right\}$ a set of region/sextant indices $\left\{p_1,\ldots,p_{N}\right\}$.
Next, we use the symmetry operation to transform the particle coordinates to a unique orthant where all training and inference is performed.
In practice, we convert them into a number in base $n$, and choose the smallest number as our representative orthant.

We emphasize that this simple, highly efficient and general technique for enforcing spatial symmetry is only viable because path integration removes the need for continuous or differentiable ansatzes.
In VMC, the discontinuities at orthant interfaces would contribute infinite kinetic energy; ad hoc ansatzes with the crystal symmetry built into them are therefore necessary \cite{guclu_interaction-induced_2008,chin_solving_2020}.
Otherwise, the symmetry breaking observed in methods formulated around the semiclassical limit \cite{filinov_wigner_2001,ghosal_incipient_2007} is difficult to reproduce within VMC, though not impossible \cite{ruggeri_nonlinear_2018,pescia_message-passing_2023}. 
Here, however, we can easily treat the entire range of interaction strengths with and without symmetry breaking, all using the same generic ansatz.

We also note that the language above should not be taken to imply a formal claim that the ground state of the system exhibits broken rotational symmetry.
Strictly speaking, there is always a symmetric ground state, and important techniques exist in the literature for obtaining expressions for it given approximate symmetry-broken states \cite{yannouleas_strongly_2002,baksmaty_rapidly_2007,yannouleas_symmetry_2007}.
In the present context, this could be effectively accomplished simply by enforcing sufficiently high-order rotational symmetry.
The signature of hidden crystal structure would then still be detectable in the pair correlation function.
However, the exact ground state becomes increasingly difficult to stabilize and observe as one approaches the classical limit, and the broken states form an effective ground state \cite{anderson_more_1972}.

In the middle and left columns of subpanels in each of the two panels of Fig.~\ref{fig:wigner}, respectively, we enforce 5-fold and 6-fold rotational symmetry on the wavefunction.
We also enforce reflection symmetry through a plane passing through the middle of each sextant.
The symmetry enforcement does not change the converged energy by more than $\sim 0.5\%$ for any given system, suggesting that the wavefunctions possessing these different densities all are close enough to the true ground state of the system that we cannot distinguish them from it without substantial further numerical effort.
Interestingly, the 5-fold states that we expect from Wigner crystallization near the classical limit appear at several interaction strengths to be closely competitive with symmetry-unbroken and 6-fold states.
This suggests that it may be possible to stabilize and observe such states experimentally, for example by suppressing the electronic density near the center of the trap, especially in the high-spin state.

\section{Discussion}
We presented an alternative to variational Monte Carlo (VMC) for approximating the ground states of many-body quantum systems.
The main idea is to combine the stochastic representation of wavefunctions (SRW) with a path integral approach to performing imaginary time propagation.
The resulting method is more general than VMC in the sense that it is meaningful and efficient to work with non-differentiable, discontinuous real-space, first-quantized ansatzes.
This enables a very inexpensive sorting-based framework for enforcing fermionic and bosonic exchange symmetries, as well as spatial symmetries.
The computational cost then increases as $\mathcal{O}\left(N\log N\right)$ with the number of particles $N$, in contrast to Slater determinant ansatzes with an $\mathcal{O}\left( N^3 \right)$ and recent Vandermonde ansatzes that reduce this to $\mathcal{O}\left( N^2 \right)$.

The discontinuous ansatzes we use have infinite energy in the variational sense, and we therefore proposed a decay-tracking approach to energy evaluation and convergence tests.
When the method works well, the final ansatz obtained after a sufficiently long time propagation should be similar to the ground state, and therefore close to a smooth function.
We showed how a variational upper bound for the ground state energy can then be obtained from it.

The time propagation steps in the algorithm suffer from a sign problem in the formal sense.
However, the stochastic noise is exponential in the size of the time step, $\Delta\tau$, rather than in the total propagation time $\tau$ as in a path integral calculation.
This means that in practice the sign problem, though still generally exponential in the number of particles $N$, can be controlled by taking smaller time steps.

Using Hooke's atom---a model often used to study electrons in 2D and 3D quantum dots---we benchmarked our path-integral-based SRW approach against exact results in the noninteracting limit, using a minimal deep neural network as an ansatz.
We then performed benchmarks against established methods from the literature, finding that the SRW is consistent with known results.
Convergence properties, computational scaling with system size, and the dependence on numerical parameters were also explored.
Despite the fact that we used a very simple and generic ansatz, we found evidence that high-dimensional wavefunctions could be represented with reasonable accuracy without an exponential increase in cost as the system becomes larger.

We then continued to investigate the interaction energy and density of the Hooke's atom at weak interaction strength, for different particle statistics, system sizes, and spin configurations.
This showcased the ability of the SRW and neural network ansatz to capture the physics of shell formation.
Focusing on the 2D, 6-fermion Hooke's atom, which has a noninteracting ground state with ``Noble''-like filled shells in both its low-spin and high-spin configurations, we plotted the density at different interaction strengths and with different spatial symmetries enforced.
This revealed that the SRW can capture the spontaneous symmetry breaking transition leading to classical Wigner molecules at large interaction strength, but also showed that in some cases low-lying states with non-trigonal symmetries may be possible to stabilize.

Looking forward, a more efficient implementation of the path integration step and more advanced machine learning architectures may enable the SRW technique to produce higher accuracy results and scale to significantly larger systems.
The choice of architecture and hyperparameters, which can be changed at each regression step in the algorithm, will certainly be important in general; this is a promising direction for future work.
An obvious extension that will benefit greatly from the enforcement of spatial symmetries is to treat atomic and molecular systems.
Unless soft pseudopotentials are used, this requires path integrals formulated so as to deal with divergent ionic potentials \cite{kleinert_path_2009}.
Another direction that should immediately provide state-of-the-art accuracy is to implement fixed-node diffusion Monte Carlo based on the smoothed SRW as an input.
As an interesting application, the methodology could potentially provide interesting new tools for studying recent experiments where Wigner crystal physics was observed in moir\'e superlattices \cite{yannouleas_quantum_2023,yannouleas_wigner-molecule_2024,li_wigner_2024}.

More speculatively, the SRW methodology presented here goes beyond VMC in being compatible not just with discontinuous ansatzes for wavefunctions, but also with probabilistic ones: i.e., where we do not know the exact value of the wavefunction at a given point, but can draw guesses for it from a distribution.
This suggests intriguing new research directions, such as---for example---using generative diffusion models to represent and progressively improve our limited knowledge about a wavefunction.
We expect the general, simple, and practical nature of the SRW method to engender a wide variety of such ideas.

\section*{Acknowledgements}
We thank Riccardo Rossi, Giuseppe Carleo, Andreas Savin, Benjamin Sorkin, and Ido Zemach for fruitful discussions and suggestions.
G.C. acknowledges support by the Israel Science Foundation (Grants No. 2902/21 and 218/19) and by the PAZY foundation (Grant No. 318/78).

\bibliography{refs.bib}
\end{document}